\documentclass[preprint2]{aastex}

\usepackage{epsfig}

\def\deg{^{\circ}}
\def\P3hat{{\mathaccent 94 P}_3}

\def\RM{{\rm RM}}
\def\RMave{{\tt RM}(\phi)}
\def\RMA{{\cal RM_{\rm A}(\phi)}}
\def\RMB{{\cal RM_{\rm B}(\phi)}}
\def\psia{{\chi_{\rm A}(\phi,\nu)}}
\def\psib{{\chi_{\rm B}(\phi,\nu)}}
\def\psiav{{\langle\chi_{\rm AB}(\nu)\rangle}}
\def\snr{{\tt S/N~}}
\def\xa{{X_{\rm A}(t,\nu)}}
\def\xb{{X_{\rm B}(t,\nu)}}
\def\pa{{\tt PA~}}
\def\pas{{\tt PA}s}

\shorttitle{Rotation Measures of Pulsars}
\shortauthors{R. Ramachandran, et al.}

\begin{document}

\title{Effect of Quasi-orthogonal Emission Modes on the Rotation
Measures of Pulsars}

\author{R. Ramachandran \& D. C. Backer} 
\affil{Department of Astronomy, University of California, 
Berkeley, CA 94720-3411, USA; \\ e-mail:
ramach,dbacker@astro.berkeley.edu}

\author{Joanna M. Rankin}
\affil{Physics Department, University of Vermont, Burlington, 
       VT 05405 USA; \\ email: joanna.rankin@uvm.edu}

\author{J. M. Weisberg \& K. E. Devine} 
\affil{Department of Physics \& Astronomy, Carleton College, 
Northfield, MN 55057; \\ email: jweisber@carleton.edu}

\begin{abstract}
We report here the discovery of a significant source of systematic
error in the rotation measure determinations of pulsars. Conventional
analysis of high sensitivity polarimetric observations of PSR B2016+28
display variation of the rotation measure of $\pm$15 rad m$^{-2}$
(around the mean value of --34.6 rad m$^{-2}$) across the pulse
profile. Analysis of single pulse data shows that this variation is an
artifact of the incoherent superposition of quasi-orthogonal
polarisation modes along with the frequency dependence of relative
strength and/or quasi-orthogonality of the modes. Quasi-orthogonal
polarization is common among pulsars, and therefore this effect needs
to be taken into account in the interpretation of pulsar rotation
measures.
\end{abstract}


\section{Introduction}
\label{sec-intro}
Pulsars play a major role in our understanding of interstellar
magnetic field structure owing to measurements of Faraday rotation of
the plane of linear polarization as a function of radio frequency
introduced by the component of the field along the sight line in the
warm interstellar medium ($B_{\parallel}$). The standard definition of
rotation measure (RM) in c.g.s. units is given by

\begin{equation}
\RM=\frac{e^3}{2\pi m_e^2c^4}\int_0^L n_e(l)\;B_{\parallel}(l)\;{\rm d}l
\label{eq:rm}
\end{equation}


\noindent
where $L$ is the distance to the pulsar from the observer, ${\rm d}l$
is the length element along the line of sight, and $n_e$ is the free
electron number density. $e$ and $m_e$ are the charge and the mass an
electron, and $c$ is the velocity of light in vacuum. The amount of
rotation experienced by the intrinsic linear polarisation position
angle of the source ($\psi_{\circ}$) at a given wavelength ($\lambda$)
is expressed as

\begin{equation}
\psi - \psi_{\circ}\;=\; \RM\;\lambda^2,
\label{eq:rmeqn}
\end{equation}

\noindent
where $\psi$ is the apparent linear polarisation position angle ({\tt
PA}) as seen by the observer. Several investigators in the past have
measured RM values of pulsars (Manchester 1972; 1974; Hamilton et
al. 1981; Hamilton \& Lyne 1987; Rand \& Lyne 1994; Qiao et al. 1995;
Weisberg et al. 2003). Time dependence in the measured RM
values\footnote{It is also important to compensate for the variable
ionospheric contribution to the measured RM values. As Manchester
(1972) states, this contribution could be as high as some 0.1 to 6 rad
m$^{-2}$, depending on the time of the day and the declination of the
pulsar.} have also been noted against some pulsars like the ones in
the Vela and Crab supernova remnants (Hamilton et al. 1977; Rankin et
al. 1988).  A positive RM means the direction of magnetic field is
towards the observer, and a negative RM means otherwise.  Using the
measured values, detailed modelling of magnetic field structure has
been carried out by several authors in the past (Thompson \& Nelson
1980; Lyne et al. 1989; Indrani \& Deshpande 1998; Han et al. 1999;
2002; Mitra et al. 2003).  All of these galactic models based on
pulsar RMs have a central assumption that the RM is completely
determined by the interstellar medium, and that the magnetosphere of
the pulsar, with all of its complexities, does not contribute
significantly.

We show in this paper that a conventional RM analysis based on average
pulse profiles leads to large variations of the RM across the pulse
%
%
profile of PSR B2016+28. The conventional analysis requires some form
of averaging to arrive at a single RM. Additional work that we are
doing has shown that this effect occurs in other pulsars as well. In
pulsars with significant RM variations across the pulse profile, there
is a source of error that may not have been considered in past
modeling of galactic magnetic fields.

%

Significant magneto-ionic propagation effects are not expected in
pulsar magnetospheres owing to the ultra-relativistic nature of the
plasma. Any large Faraday rotation within the emission region of the
magnetosphere would lead to severe depolarisation across our band. The
fact that pulsar radiation is highly polarised therefore shows that
there is no significant Faraday rotation within the emission
region. We are led then to look more closely at the data to determine
the nature of the RM variations across the pulse.

We show in this work an improved method for RM determination when
sufficient signal-to-noise ratio ({\tt S/N}) allows detection of the
pulsed radiation in single pulses. In this case, the orthogonal modes
of polarization (e.g., Backer \& Rankin 1980) can be identified, and
the RM can be determined from each mode independently. If the modes
were strictly orthogonal AND there was no jitter AND there was no
frequency dependence of these properties, then this procedure would
not be necessary. However, the \pas ~(position angles) at each
longitude do display jitter and a slight degree of non-orthogonality
(Gil, Snakowski \& Stinebring 1991; Gil et al. 1992). We show in this
work that the B2016+28 emission has a slight frequency dependence in
the relative strengths and (or) amounts of non-orthogonality. This is
the source of the pulse-longitude dependence of the RM in the
conventional analysis. One does not have to appeal to strong, and
unexpected, magneto-ionic propagation effects.

Most pulsars for which any amount of single pulse study exists show
signs of orthogonal modes in their emission. The problem of origin of
these modes has been addressed by several investigators (e.g., Melrose
1979; Petrova \& Lyubarskii 2000; Radhakrishnan \& Deshpande 2001;
Petrova 2001). As they describe, the orthogonal mode could arise from
the partial conversion of an original mode (ordinary) to the other
(extra-ordinary) due to propagation in the magnetosphere of the
pulsar. Even the cause for the slight non-orthogonality has been
addressed by investigators like Petrova (2001).

In this paper, after briefly describing the details of our
observations in Sec.~\ref{sec-observation}, we summarize the
conventional approach to RM determination in
Sec.~\ref{sec-conventional}. Here we describe the aforementioned RM
variations as a function of pulse longitude for PSR B2016+28. In
Sec~\ref{sec-intrinsic} \& \ref{sec-nonorthog} we show how
quasi-orthogonal modes in pulsars can cause severe artifacts in RM
estimations, and in particular how the apparent variations in the RM
of PSR B2016+28 can be produced.  We conclude our analysis with a
detailed discussion in Sec.~\ref{sec-discussion}.  The symbols for
variables used in this paper are defined in the appendix.

\begin{figure*}
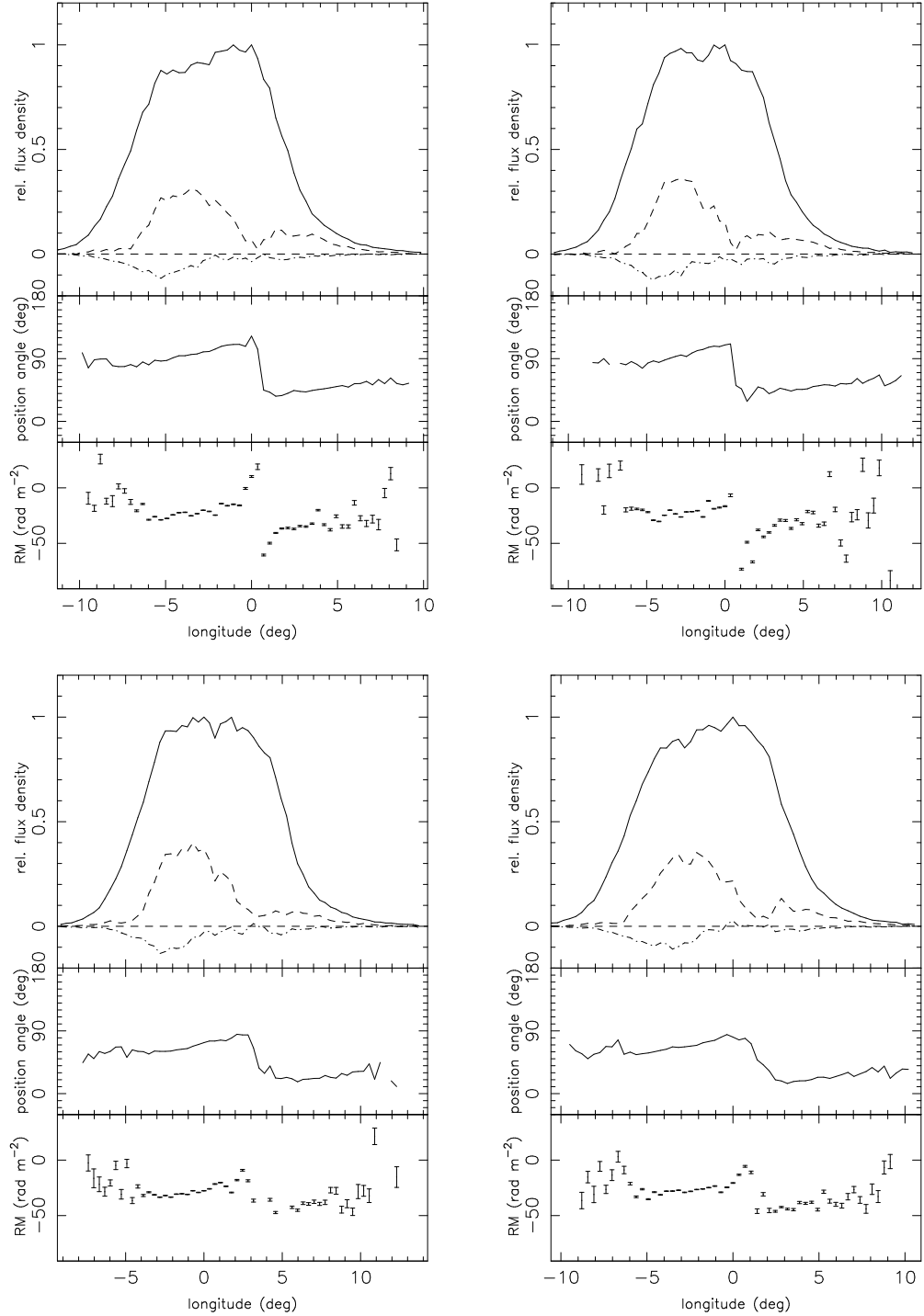

\begin{center}
\epsfig{file=fig1a.ps,width=9.5cm,angle=-90} \\

\vspace*{0.2cm}

\epsfig{file=fig1b.ps,width=9.5cm,angle=-90}
\end{center}
\caption{(a) -- (d) : Results from four different average-pulse data
sets. Top panel gives the average profile in Stokes I parameter (solid
line), linearly polarised power (`dash' line) and circularly polarised
power (`dot-dash' line). {\bf Middle:} Linear polarisation position
angle determined from the average profile. {\bf Bottom:} $\RMave$ vs
$\phi$. See text for details.}
\label{fig:more2016}
\end{figure*}

\section{Observations}
\label{sec-observation}
The average pulse observations were performed at the Arecibo
Observatory at 430 MHz in May and December 1992 with a band width of 5
MHz. Average pulse profiles were produced by integrating the signal in
each frequency channel for 120 seconds. The observation setup has been
described in detail in Weisberg et al. (2003).

The single-pulse observations were carried out at Arecibo Observatory
at 430-MHz centre frequency in a single session on October 1992. With
a special program for gating the 40-MHz correlator, auto-correlation
and cross-correlation functions of the right-hand and left-hand
polarisation-channel voltages were recorded with 32 correlation lags.
The bandwidth of the observation was 10 MHz.  The correlation vectors
were averaged to a time resolution of 506 $\mu$sec. The correlation
data were three-level corrected after scaling, and Fourier transformed
to produce Stokes parameter spectra. A detailed calibration procedure
was adopted to correct for instrumental effects, delays and
interstellar dispersion. These observations as well as calibration
procedures have been described in Rankin \& Rathnasree (1995). The
resulting polarised pulse sequence had 32 channels (although channel 1
lacked Stokes U and thus was unusable) with 3043 pulses (1600 sec),
and was gated to include a longitude range of some 45$\deg$. A
constant RM value of --34.6 rad m$^{-2}$ (earlier measurement in the
literature) was also subtracted from the data.

\begin{figure*}[t]
\begin{center}
\epsfig{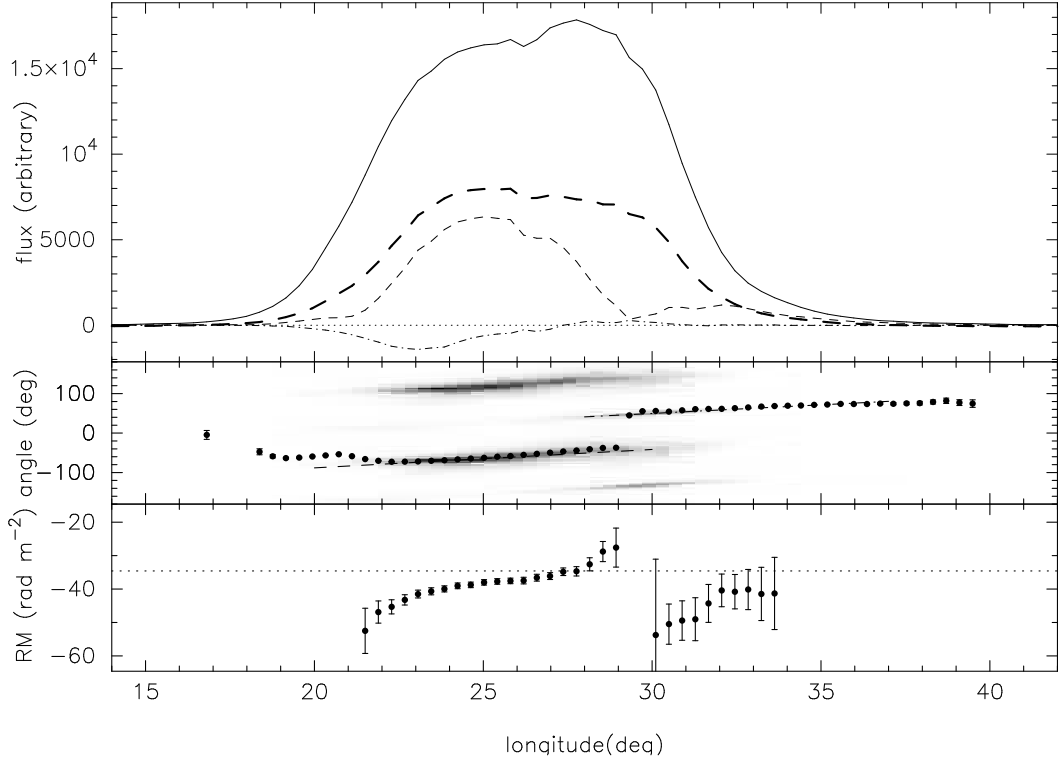}
\end{center}
\caption[]{$\RMave$ vs $\phi$. {\bf Top : } Average pulse profile in
Stokes-I is given as solid line. `Thin' and `thick' dash lines are
linear polarisation power computed from average pulses and from
single-pulses and averaged over all pulses, respectively. `Thin'
dot-dash line gives circular polarisation power (Stokes V) computed
from average pulses. A horizontal dotted line is drawn to indicate the
zero-level in the Y-axis. {\bf Middle : } The dots show position angle
estimated from the average profile. The gray scale gives the
probability density function of position angles estimated from
single-pulses. {\bf Bottom : } $\RMave$ estimated from average
pulse. The value of $\langle {\bf RM}\rangle$ computed from this is
--38.3(3) rad m$^{-2}$.}
\label{fig:RMave}
\end{figure*}
\section{Conventional RM measurements}
\label{sec-conventional}
As individual pulses from pulsars are in general faint, one typically
generates an ``average'' pulse profile by folding the time series in
each frequency channel and Stokes parameter synchronously with the
Doppler-shifted apparent period. An important property of radio
emission from pulsars is that the position angle of linear
polarisation changes as a function of pulse longitude (Radhakrishnan
\& Cooke 1969). Owing to this property, \RM ~measurements of pulsars
have always been more complex than for any other polarised radio
source in the sky. \pa measurements using a given longitude bin of
folded profiles and all frequency channels are used to fit for the
rotation measure [$\RMave$] on the basis of Eqn.~\ref{eq:rmeqn}. This
is repeated for each longitude bin, and the \RM ~from all these bins
are averaged to estimate the overall value, $\langle{\bf RM}\rangle$,
for the pulsar. Of course, the underlying assumption here is that the
value measured in all the longitude bins are entirely introduced by
only the interstellar medium, and hence there is no
longitude-dependent Rotation Measure introduced by the pulsar itself.
Errors in the $\RMave$ can be simply propagated using the \snr of each
individual estimate and an assumption of independence of the
estimates.

\begin{figure*}[t]
\begin{center}
\epsfig{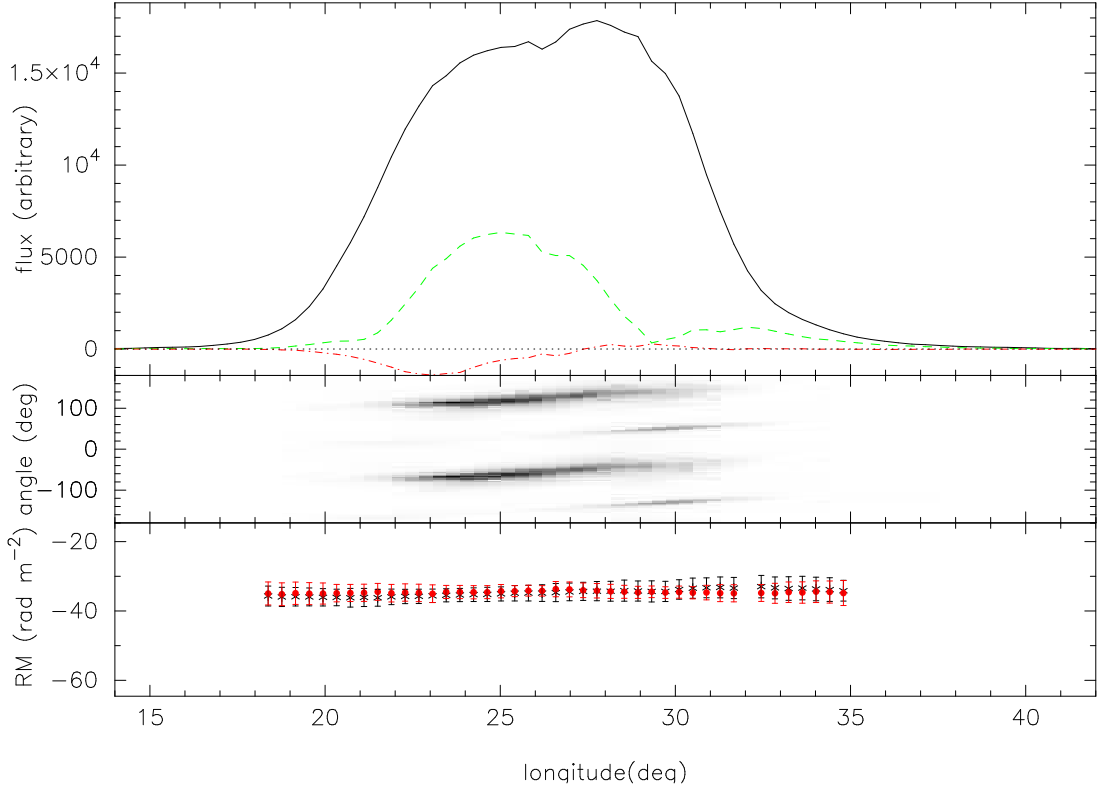}
\end{center}
\caption[]{Rotation Measure of the two polarisation modes [$\RMA$ \&
$\RMB$] as a function of pulse longitude. For comparison, we have
given average profiles in total power (solid), linear (dash) and
circular (dotted) polarisation in the top panel, and position angle
PDF in the middle panel as gray scale. The bottom panel gives the
Rotation Measure of the two modes, $\RMA$ and $\RMB$. Effective values
of RM of the two modes are --35.0(4) \& --34.5(3) rad m$^{-2}$.}
\label{fig:RMmodes}
\end{figure*}

In our analysis, as the first step, we have attempted to compute
$\RMave$ values as a function of pulse longitude. As mentioned above,
this has been estimated from the average profiles constructed in each
of the 31 frequency channels in our sequence.  Fig.~\ref{fig:more2016}
\& ~\ref{fig:RMave} summarise the results. Fig.~\ref{fig:more2016}
shows the results from four different scans from our, each of 120 sec
integration. In the top panel, we give average pulse profile in Stokes
I parameter (solid line), average linearly polarised power (`dash'
line), and average circularly polarised power (`dot-dash' line). The
middle panel shows linear polarisation position angle, and the bottom
panel shows $\RMave$.

In the top panel of Fig.~\ref{fig:RMave}, the solid line gives the
average pulse profile in total power and the `thin' dash line gives
the average profile in linear polarisation computed from average
profiles in all Stokes parameters. This is to be compared with the
`thick' dash line, which gives linear power computed from
single-pulses and averaged over all pulses. As we can see, the `thin'
dash line shows a smaller degree of linear polarisation owing to
incoherent superposition of the polarisation modes.
%
Then the `thin' dot-dash line indicates power in circular polarisation
computed from average profiles in Stokes-V parameter. Power less than
zero just means that it is left-circularly polarised.

The dots in the middle panel give the linear polarisation \pa curve as
defined by the average profile, and the gray scale gives the
probability density function (PDF) of the position angle computed from
all the single-pulses. While computing this PDF, we have weighted the
values with the square of their S/N ratio of the polarised flux as
defined by $\sqrt{Q^2+U^2}/\sigma_{\rm sys}$ (where $Q$ \& $U$ are two
of the Stokes parameters, and $\sigma_{\rm sys}$ is the
root-mean-square value of the system noise flux\footnote{As the
telescope gain of Arecibo Telescope is significantly high, variation
of system noise with the strength of the pulsar was explicitly taken
into account with the help of calibrated average pulse
profile.}). Although the \pa range outside $\pm 90\deg$ is
redundant, we have chosen a range of $\pm 180\deg$ for clarity. 

In the bottom panel of Fig.~\ref{fig:RMave}, we give the measured
values of $\RMave$ as a function of pulse longitude. The dotted line
in the bottom panel shows the earlier measurement of RM for this
pulsar (Manchester 1972), --34.6 rad m$^{-2}$. As we mentioned in
Section \ref{sec-observation}, the data that we used for our analysis
already had a constant RM value of --34.6 rad m$^{-2}$ removed from
it. But to be consistent, we have added this offset in the plot
(dotted line).  Effective RM from our measurement comes to --38.3(3)
rad m$^{-2}$, which is significantly different from the earlier
measurement of Manchester, as well as of Weisberg et al. (2003), which
is $-27.3\pm 2.1$ rad m$^{-2}$.

There are two aspects of Fig.~\ref{fig:more2016} \& \ref{fig:RMave}
that needs to be discussed here. The measured values of $\RMave$ as a
function of pulse longitude display significant systematic
variations. If this effect is true, then it has a very fundamental
significance, as one does not expect the interstellar medium to
distinguish between one pulse longitude to the other. Therefore, this
RM difference must be due to intrinsic propagation effects in the
pulsar magnetosphere. In fact, this would become the first ever direct
evidence for propagation effects in the pulsar magnetosphere. However,
as we will show in the following section, this is due to artifacts
introduced by superposition of the two more or less orthogonal
emission modes seen in this pulsar.

Secondly, in Fig.~\ref{fig:more2016}, the overall behaviour of
$\RMave$ is similar in all plots. As described in
Sec.~\ref{sec-observation}, these data sets were obtained in May and
December 1992. The ``anti-symmetric'' behaviour of $\RMave$ (with
respect to longitude value of zero in Fig.~\ref{fig:more2016}) seems
to be stable over a time scale of several months. However, although
the overall behaviour of $\RMave$ is stable, it is quite clear that
the exact value of $\RMave$ is not the same in all the panels of
Fig.~\ref{fig:more2016}. We will return to this later in
Sec.~\ref{sec-discussion}.

RM values varying as a function of pulse longitude is not unique to
this pulsar alone. It turns out that this has been observed in other
pulsars like PSR B0329+54 (Mitra, Johnston \& Kramer, private
communication).

\section{Intrinsic $\RMave$ variations?}
\label{sec-intrinsic}
If the two polarisation modes are strictly orthogonal, then they have
no effect on conventional RM measurements. When we compute the average
profile in each frequency channel, all that we are doing is
``incoherently'' superposing the radiation from the two orthogonal
modes. This would mean that the net degree of polarisation is the
difference of their degrees of polarisation, and the \pa direction is
the same as that of the dominant mode. Provided adequate resultant
polarised signal, we can still estimate the RM value. If the modes are
not strictly orthogonal, then the net \pa of the average is the result
of a vector summation.  This too is of no consequence to RM
determination as long as the relative strength of the two modes and
their individual mean \pa remain constant as a function of radio
frequency. 

There have been only a few quantitative studies of the frequency
dependence of the orthogonal modes (Karastergiou 2002; 2003). In
general the degree of polarization of this pulsar decreases
significantly with increasing frequency (Gould \& Lyne 1998). Also, as
the observations of Gould \& Lyne show, the mode-dominance transition
point (longitude of $\sim$ 29$^{\circ}$ in Fig.~\ref{fig:RMave}) moves
to earlier longitudes with increasing radio frequency.  This indicates
that the relative strengths of the modes {\it do not} remain constant
as a function of frequency. In other words, the first half of the
profile, which is predominantly polarised with mode-A (represented by
a dash line) has a steeper spectral index when compared to the second
half that is predominantly polarised with mode-B (dot-dash line).

In the case of PSR B2016+28, it is also clear that the modes are not
exactly orthogonal. The best fit to the two gray-scale tracks show
that the \pa and the slope at $30\deg$ longitude are : $-42\deg .8
(2)$ and $5.04(1) \deg/\deg$ for mode-A (represented by the dash line)
and $49\deg .8(1)$ and $4.37(1) ^{\circ}/^{\circ}$ for mode-B
(dot-dash line). These two slope values are significantly different
from each other. This \pa non-orthogonality is known in the literature
(Gil, Snakowski \& Stinebring 1991; Gil et al. 1992), and has recently
been studied in detail by McKinnon (2003). In fact, it has been shown
by McKinnon that this is fairly common among several pulsars
(e.g. PSRs B0950+08, B1929+10, B2020+28).

\section{RM determination in presence of quasi-orthogonal modes}
\label{sec-nonorthog}
Although the fractional polarisation seen in the average pulse
profiles are typically a few tens of percent, individual pulses can
exhibit even higher degree of polarisation, some times nearly reaching
hundred percent (Stinebring et al. 1984; Rankin et al. 1989;
Ramachandran et al. 2002; Rankin \& Ramachandran 2003). Given the
presence of the quasi-orthogonal modes and possible depolarisation due
to their superposition, quite independent of any possible artifacts,
it makes sense for us to determine the RM with individual pulses
rather than average pulses.

As described in Section \ref{sec-observation}, our single pulse data
set of PSR B2016+28 consists of 3843 pulses, gated to represent a
longitude range of about 45$\deg$ in each pulse. As the first step, in
each frequency channel, we produced a \pa distribution for every
longitude. In doing this, we weighted each sample by the square of
\snr of the polarised flux. At a given longitude, due to the presence
of the two modes, one expects two significant peaks in this
distribution separated by almost 90$\deg$.

As the second step, by a weighted mean, we found the centroid of the
two peaks [$\psia$ and $\psib$], both being functions of pulse
longitude and radio frequency. We thus determined the mean \pa values
of each mode.

\begin{figure}[t]
\begin{center}
\epsfig{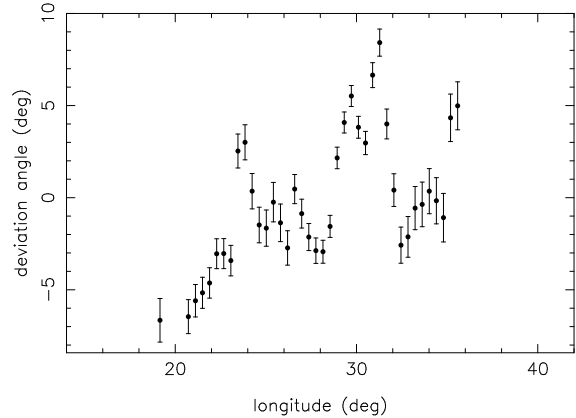}
\end{center}
\caption[]{Average deviation angle (Y-axis) from non-orthogonality of
the two polarisation modes as a function of pulse longitude
(X-axis). See text for details.}
\label{fig:nonortho}
\end{figure}

Once the \pas ~are determined, then it is straight forward to fit for
rotation measure values of the two modes, $\RMA$ and $\RMB$, as a
function of pulse longitude. These two functions determined for PSR
B2016+28 are given in Fig.~\ref{fig:RMmodes}. Comparing this to the
bottom panel of Fig.~\ref{fig:RMave}, it is obvious that the
deviations of RM values observed for the two modes is far less than
what is determined with the average profiles. The effective RM values
of the two modes turn out to be $-35.0\pm 0.4$ and $-34.5\pm 0.3$ rad
m$^{-2}$, respectively. These values are very close to the earlier
measurement of $-34.6\pm 1.4$ rad m$^{-2}$ (Manchester 1972).
%

As emphasised in Section~\ref{sec-intrinsic}, the presence of the two
modes itself, even if they are not orthogonal, should not introduce
any artifact in RM measurements, as long as there is no frequency
dependence of the relative strength and the intrinsic \pas of the two
modes. However, in our analysis of PSR B2016+28, the value of $\RMave$
is significantly different from the values of $\RMA$ and $\RMB$, which
clearly indicates that the incoherent superposition of the modes while
generating average profiles (as a function of frequency) has
introduced a frequency dependence of the resultant \pa, which
manifests as {\it extra} RM.

To investigate this subtle effect in detail, with the help of the
above described procedure, we produced Fig.~\ref{fig:nonortho}, that
shows the non-orthogonality of the two modes as a function of pulse
longitude. We have plotted the quantity
\begin{equation}
\Delta\psi(\phi) = \frac{\pi}{2} - \left \langle
[\psib - \psia] \right \rangle_{\nu}.
\end{equation}
\noindent
against pulse longitude ($\phi$). As we can see, the deviations are
not uniform across the pulse profile, and the variations are
statistically significant.

Let $\xa$ and $\xb$ be the ``instantaneously'' randomly varying
linearly polarised intensities of the two quasi-orthogonal modes. Let
$\zeta(\nu)$ be the angle between the two vectors in the Stokes Q--U
space. Assuming that the \pa of the primary mode is zero, we can write
the two Stokes parameters $Q$ and $U$ as (see also McKinnon 2003)
\begin{eqnarray}
Q(t,\nu) &=& \xa +\xb\;\cos\zeta(\nu)+X_{\rm Q}(t,\nu)\nonumber \\
U(t,\nu) &=& \xb\;\sin\zeta(\nu)+X_{\rm U}(t,\nu)
\end{eqnarray}
If $\zeta(\nu)$ is independent of $\xa$, then the
\pa as a function of frequency with average
values of $Q$ and $U$ is
\begin{eqnarray}
\psiav &=& \frac{1}{2}{\rm tan^{-1}}\left[ \frac{\langle U(t,\nu)
\rangle}{\langle Q(t,\nu) \rangle} \right] \nonumber\\ &=&
\frac{1}{2} {\rm tan^{-1}}\left[ \frac{\sin\zeta(\nu)}{F_{\rm AB}(\nu) + \cos
\zeta(\nu)}\right]
\label{eq:psiav}
\end{eqnarray}
\noindent
where the angular brackets indicate time-averaging, and $F_{\rm
AB}(\nu) = \langle\xa\rangle / \langle\xb\rangle $. $X_{\rm
Q}(t,\nu)$ and $X_{\rm U}(t,\nu)$ are the system noise strengths
in $Q$ and $U$ as a function of time and frequency. As we can see, in
principle, frequency dependence in $\zeta(\nu)$ or $F_{\rm AB}(\nu)$
can introduce a frequency dependent $\psiav$, which can corrupt our
rotation measure determination.

The dependence of $\psiav$ on frequency is different from that of the
RM. Therefore, in principle, the RM measured at different frequency
ranges should be different. To check this, we divided our frequency
band into two parts. Indeed, the value of $\RMave$ was different in
the two halves. In the first half of the band, the measured value of
RM was $-32.5 (9)$ rad m$^{-2}$, whereas in the other it was $-40.5
(4)$ rad m$^{-2}$. The overall value of $\langle{\bf RM}\rangle$ was
$-38.3 (6)$ rad m$^{-2}$.

\begin{figure*}
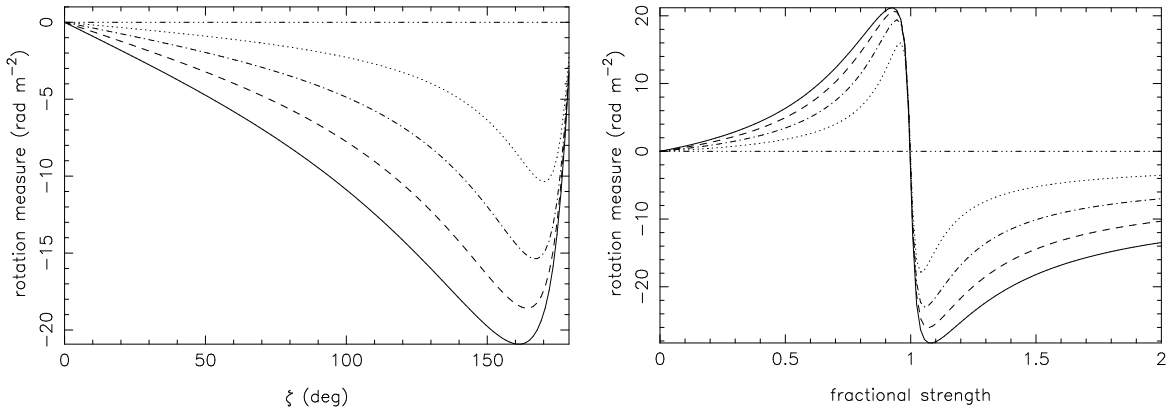

\begin{center}
\begin{tabular}{@{}lr}
{\mbox{\epsfig{file=f5a.ps,height=7.5cm,angle=-90}}} & 
{\mbox{\epsfig{file=f5b.ps,height=7.5cm,angle=-90}}}\\
\end{tabular}
\end{center}
\caption[]{Effective RM introduced by non-orthogonal mode-mixing
(Y-axis). Left panel shows RM as a function of $\zeta(\nu)$. For a
constant $\zeta(\nu)$ value, the five curves, {\it solid}, {\it dash},
{\it dot-dash}, {\it dotted} and {\it dot-dot-dot-dash}, correspond to
variations of fractional strengths of the two modes ($F_{\rm
AB}(\nu)$) from the lower to the upper end of the frequency band of
0.3--0.9, 0.45--0.9, 0.6--0.9, 0.75--0.9 and a constant 0.9,
respectively. In the right panel, the five curves correspond to the
variation of $\zeta(\nu)$ of 160$\deg$--180$\deg$,
165$\deg$--180$\deg$, 170$\deg$--180$\deg$, 175$\deg$--180$\deg$,
constant value of 180$\deg$, respectively. See text for details.}
\label{fig:th_frac_rm}
\end{figure*}

The error introduced by this effect depends on the bandwidth, centre
frequency, frequency dependences of $\zeta(\nu)$ and $F_{\rm
AB}(\nu)$. Fig.~\ref{fig:th_frac_rm} succinctly summarises this
effect. This theoretically generated plot corresponds to a centre
frequency of 430 MHz and a band width of 10 MHz, which is divided into
128 frequency channels. In the left panel, we give $\zeta(\nu)$ in the
X-axis and effective RM introduced by this quasi-orthogonal
mode-mixing in the Y-axis. The five curves, {\it solid}, {\it dash},
{\it dot-dash}, {\it dotted} and {\it dot-dot-dot-dash}, correspond to
variations of fractional strength of the two modes [$F_{\rm AB}(\nu)$]
from the lower to the upper end of the frequency band of 0.3--0.9,
0.45--0.9, 0.6--0.9, 0.75--0.9 and a constant 0.9, respectively. As is
seen, when the relative strength remains constant, then the RM
introduced by this effect is zero. However, when the relative strength
varies as a function of frequency, then the spurious RM introduced in
the measurement could be significant.

In the right panel of Fig.~\ref{fig:th_frac_rm}, we have addressed the
other possibility that for a constant fractional strength across the
frequency band, $\zeta(\nu)$ varies. The five curves, {\it solid},
{\it dash}, {\it dot-dash}, {\it dotted} and {\it dot-dot-dot-dash},
correspond to $\zeta(\nu)$ varying in the ranges of
160$\deg$--180$\deg$, 165$\deg$--180$\deg$, 170$\deg$--180$\deg$,
175$\deg$--180$\deg$, and a constant value of 180$\deg$,
respectively. Here, the RM value changes sign at $F_{\rm AB}(\nu) =
1$, and this is exactly what we see in Fig.~\ref{fig:RMave}. The
inferred value of $\RMave$ has an anti-symmetry with respect to
$\phi\sim 30\deg$, where one mode is stronger to the right and the
other mode to the left. It is worth noting here that in
Fig.~\ref{fig:th_frac_rm}, at the limit of $F_{\rm AB}(\nu)
\longrightarrow \infty$, the RM introduced asymptotically reduces to
zero.

\section{Discussion}
\label{sec-discussion}
In Fig.~\ref{fig:more2016} \& \ref{fig:RMave}, one of the intriguing
things is that the value $\RMave$ does not remain constant between
various epochs of observation. Perhaps one of the reasons could be a
combination of interstellar scintillation and the effect that we have
addressed in this paper. Although scintillation is not expected to
have any effect on the amount of Faraday rotation introduced in the
interstellar medium, when the total integration time is short ($\sim$
a few minutes), the effective centre frequency and bandwidth is
expected to vary. Combined with the effect that we have described, the
RM value inferred in principle can be different at different
epochs. Moreover, as described in Sec.~\ref{sec-observation}, each of
the four panels (a to d) in Fig.~\ref{fig:more2016} has been produced
with 120 second ($\sim$200 pulses) long scans. Pulsars are known to
exhibit stable average-pulse profiles only after integrating a few
thousands of pulses. As we can see from the four panels, average
profiles, and even polarisation-sweep curves, are not identical
between all the panels. This may also introduce time-dependence in RM
values.

As mentioned in Sec.~\ref{sec-nonorthog}, the exact frequency
dependence of $F_{\rm AB}(\nu)$ and $\zeta(\nu)$ are not known. In
general, there is no reason to conclude that the frequence dependence
of $\psiav$ introduced by this effect is the same as that of the
RM. What we have attempted in Fig.~\ref{fig:th_frac_rm} is to estimate
the effective RM introduced by this effect. A thorough investigation
of the nature of the emission modes and their frequency dependence is
beyond the scope of this publication, and we plan to consider these
matters in a subsequent report.

As radiation from pulsars is highly polarised and as distance
estimates to pulsars are more reliable than the other polarised
sources (e.g. supernova remnants), pulsar RM determinations have been
used extensively in probing the magnetic field structure of the
Galaxy. RM is an important tool to model the long range as well as the
turbulent component of the magnetic field in the Galaxy. If the effect
that we have discussed in this paper is common, then it will have
serious consequences on the existing RM measurements of pulsars, and
thereby on the models of magnetic field structure in the Galaxy. Apart
from our detailed analysis of PSR B2016+28 presented here, in the
sample of Weisberg et al. (2003), we looked at pulsars B0301+19,
B0525+21, B0626+24, B1929+10 and B2020+28. These all showed systematic
variation of the RM across the pulse profile. For instance, Weisberg
et al. measure RM values for these pulsars as --5.7 ($\pm$10), --39.2
($\pm$10), 69.5 ($\pm$10), --5.9 ($\pm$5), --73.7 ($\pm$25) rad
m$^{-2}$, respectively. The values given in parenthesis are the
approximate magnitude of variations seen around the mean value across
the pulse profile. As we can see clearly, these variations are
significant. Especially, when the magnitude of RM is small, these
variations can cause significant bias.

The presence of orthogonal modes is common among pulsars. Among
pulsars for which any amount of detailed single-pulse study has been
done so far, it is clear that a great majority of them exhibit
orthogonal modes. In particular, it is almost impossible to find a
``conal'' pulsar that does not show evidence of orthogonal mode
emission associated with its conal components (Rankin \& Ramachandran
2003). It is also possible that these two modes have significantly
different spectral indices. For instance, there is clear evidence to
show that the degree of polarisation in the average-pulse profiles
decreases with increasing frequency. This can be understood easily by
having one of the modes dominating in strength at lower frequencies,
and the two modes having roughly comparable strengths at very high
frequencies. As the earlier studies (Stinebring et al. 1984; Gil,
Snakowski \& Stinebring 1991; Gil et al. 1992; McKinnon 2003) have
shown, these modes are slightly non-orthogonal, and this
non-orthogonality of the two modes is a wide-spread
phenomenon. 

As we have shown in this paper, the only way to eliminate this
artifact is to determine RM values from single pulses. This, of
course, is a very challenging task, owing to signal-to-noise ratio
considerations. For weaker pulsars for which we cannot obtain good
single pulse data, it is impossible to separate the two modes to
unambiguously determine RM values. Therefore, weaker pulsars are bound
to suffer from this artifact, and there is no obvious way of
correcting it.

On the Galactic scale, whether or not this effect will make serious
changes to the magnetic field model, remains to be seen. A project to
determine correct RM values for several other pulsars is underway, and
will be presented in a subsequent publication. It is conceivable that
this artifact will be most prominant among pulsars with small RM
magnitudes ($\le$50 rad m$^{-2}$ or so). A thorough analysis to check
the validity of the already existing RM values, and the effect on the
Galactic magnetic field structure is much needed.

\section{Conclusion}
Our major conclusions from this work can be summarised as follows:
\begin{itemize}
\item We find that the rotation measures determined for PSR B2016+28
as a function of pulse longitude varies significantly. This seems to
be the case for five other pulsars, namely PSRs B0301+19, B0525+21,
B0626+24, B1929+10 and B2020+28.
\item This effect is an artifact introduced by the frequency
dependence of relative strengths of the two modes as well as the
amount of non-orthogonality, and it is not intrinsic to the pulsar
magnetosphere. We show that if we estimate rotation measure for the
two modes separately, this effect can be removed. The technique to
remove this artifact invariably involves analysis with single
pulses. Hence, cannot be carried out for fainter objects. There is no
other obvious way of compensating for this artifact for these fainter
objects.
\item As the amount of RM spuriously introduced can be as high as a
few tens of units, several measurements of Rotation Measures of
pulsars in the literature may be in error, and needs revision.
\end{itemize}

\acknowledgements We thank Jon Arons, Avinash Deshpande, Aris
Karastergiou, Simon Johnston, Michael Kramer \& Dipanjan Mitra for
useful comments on the manuscript. Special thanks are due to Dipanjan
Mitra, Simon Johnston and Michael Kramer for sharing their results
prior to publication. We also thank our anonymous referee for his/her
valuable comments.

\appendix

\section*{DEFINITIONS}
\begin{center}
\begin{tabular}{rl} \\
  & \\
{\bf Variable} & {\bf Definition} \\
  & \\
$\phi$        & pulse longitude \\
{\tt PA}      & linear polarisation position angle \\
RM            & Rotation Measure \\ 
$\RMave$      & Rotation Measure at a given longitude $\phi$ 
                measured with average pulse profiles. \\
$\RMA$        & Rotation Measure of polarisation mode-A \\
$\RMB$        & Rotation Measure of polarisation mode-B \\
{\tt SNR}     & signal-to-noise ratio \\
$\psia$, $\psib$ & Weighted mean values of linear polarisation 
                   position angles of \\
              & Mode-A and Mode-B \\
$\psiav$      & linear polarisation position angle at a given frequency after \\
              & incoherently superposing the two (non-)orthogonal 
                modes. \\
$\xa$, $\xb$  & linear polarisation intensities of the two modes \\
              & as a function of time and frequency. \\
$X_{\rm N}(t,\nu)$   & system noise power as a function of time and frequency.\\
$\langle {\bf RM}\rangle$ & Net RM value computed from average 
                pulse profile after averaging \\
              & over all pulse longitudes. \\
\end{tabular}
\end{center}
\end{document}